\documentstyle[12pt]{article}
\def\PRL #1 #2 #3{{\sl Phys. Rev. Lett.} {\bf#1} (#2) #3}
\def\NPB #1 #2 #3{{\sl Nucl. Phys.} {\bf B #1} (#2) #3}
\def\NPBFS #1 #2 #3 #4{{\sl Nucl. Phys.} {\bf B #2} [FS#1] (#3) #4}
\def\CMP #1 #2 #3{{\sl Commun. Math. Phys.} {\bf #1} (#2) #3}
\def\PRD #1 #2 #3{{\sl Phys. Rev.} {\bf D #1} (#2) #3}
\def\PLA #1 #2 #3{{\sl Phys. Lett.} {\bf A #1} (#2) #3}
\def\PLB #1 #2 #3{{\sl Phys. Lett.} {\bf B #1} (#2) #3}
\def\JMP #1 #2 #3{{\sl J. Math. Phys.} {\bf #1} (#2) #3}
\def\PTP #1 #2 #3{{\sl Prog. Theor. Phys.} {\bf #1} (#2) #3}
\def\SPTP #1 #2 #3{{\sl Suppl. Prog. Theor. Phys.} {\bf #1} (#2) #3}
\def\AoP #1 #2 #3{{\sl Ann. of Phys.} {\bf #1} (#2) #3}
\def\PNAS #1 #2 #3{{\sl Proc. Natl. Acad. Sci. USA} {\bf #1} (#2) #3}
\def\RMP #1 #2 #3{{\sl Rev. Mod. Phys.} {\bf #1} (#2) #3}
\def\PR #1 #2 #3{{\sl Phys. Reports} {\bf #1} (#2) #3}
\def\AoM #1 #2 #3{{\sl Ann. of Math.} {\bf #1} (#2) #3}
\def\UMN #1 #2 #3{{\sl Usp. Mat. Nauk} {\bf #1} (#2) #3}
\def\FAP #1 #2 #3{{\sl Funkt. Anal. Prilozheniya} {\bf #1} (#2) #3}
\def\FAaIA #1 #2 #3{{\sl Functional Analysis and Its Application} {\bf
#1} (#2) #3}
\def\BAMS #1 #2 #3{{\sl Bull. Am. Math. Soc.} {\bf #1} (#2)
#3} \def\TAMS #1 #2 #3{{\sl Trans. Am. Math. Soc.} {\bf #1} (#2) #3}
\def\InvM #1 #2 #3{{\sl Invent. Math.} {\bf #1} (#2) #3}
\def\LMP #1 #2 #3{{\sl Letters in Math. Phys.} {\bf #1} (#2) #3}
\def\IJMPA #1 #2 #3{{\sl Int. J. Mod. Phys.} {\bf A #1} (#2) #3}
\def\AdM #1 #2 #3{{\sl Advances in Math.} {\bf #1} (#2) #3}
\def\RMaP #1 #2 #3{{\sl Reports on Math. Phys.} {\bf #1} (#2) #3}
\def\IJM #1 #2 #3{{\sl Ill. J. Math.} {\bf #1} (#2) #3}
\def\APP #1 #2 #3{{\sl Acta Phys. Polon.} {\bf #1} (#2) #3}
\def\TMP #1 #2 #3{{\sl Theor. Mat. Phys.} {\bf #1} (#2) #3}
\def\JPA #1 #2 #3{{\sl J. Physics} {\bf A#1} (#2) #3}
\def\JSM #1 #2 #3{{\sl J. Soviet Math.} {\bf #1} (#2) #3}
\def\MPLA #1 #2 #3{{\sl Mod. Phys. Lett.} {\bf A #1} (#2) #3}
\def\JETP #1 #2 #3{{\sl Sov. Phys. JETP} {\bf #1} (#2) #3}
\def\JETPL #1 #2 #3{{\sl  Sov. Phys. JETP Lett.} {\bf #1} (#2) #3}
\def\PHSA #1 #2 #3{{\sl Physica} {\bf A #1} (#2) #3}
\def\CQG #1 #2 #3{{\sl Class. Quantum Grav.} {\bf #1} (#2) #3}
\def\SJNP #1 #2 #3{{\sl Sov. J. Nucl. Phys. (Yadern.Fiz.)} {\bf #1} (#2) #3}
\def\a{\alpha}\def\b{\beta}\def\g{\gamma}\def\e{\epsilon}

\def\Th{\Theta}\def\Om{\Omega}

\newcommand{\p}[1]{(\ref{#1})}
\begin{document}
\renewcommand{\thefootnote}{\fnsymbol{footnote}}
\thispagestyle{empty}
\begin{flushright}
Preprint DFPD 96/TH/36\\
hep-th/9606147\\
June 1996
\end{flushright}

\medskip
\begin{center}
{\large\bf
Superstrings in the geometrical approach and constrained super--WZNW
models}\footnote{Based on talks given at the INTAS Network Workshop 
on ``Fundamental Problems in Classical, Quantum and String Gravity"
(Torino, 2-3 May, 1996), The International Conference on ``Problems of 
Quantum Field Theory" (Alushta, 13-18 May, 1996) and 
The Second International Sakharov Conference on Physics 
(Moscow, 20-24 May, 1996).}

\medskip
Dedicated to the memory of Dmitrij V. Volkov and Victor I. Ogievetskii

\vspace{0.7cm}
{\bf Dmitri Sorokin}

{\it 
INFN, Sezione di Padova, via Marzolo, 8, 35131 Padova, Italia
and \\
Kharkov Institute of Physics and Technology, Kharkov, Ukraine}

\bigskip
\bigskip
{\bf Abstract}
\end{center}
By use of geometrical methods of surface theory we demonstrate
links of Green-Schwarz superstring dynamics with supersymmetric 
exactly-solvable nonlinear systems and super-WZNW models reduced in an 
appropriate way.

\renewcommand{\thefootnote}{\arabic{footnote}}
\bigskip
\bigskip
This talk is based on recent results obtained in collaboration with I.
Bandos, P. Pasti, M. Tonin and D. Volkov \cite{bpstv,lio}, and F. Toppan
\cite{wz} in studying classical dynamics of supersymmetric extended objects
in a geometrical (twistor--like) approach (see \cite{bpstv} for references),
which is based on following principles:

\noindent
i) the fermionic $\kappa$--symmetry of the Green--Schwarz formulation of
super--p--branes is a manifestation of superdiffeomorphisms of 
worldvolume
supersurfaces  of the super--p--branes \cite{stv}. 
(This solves the problem of infinite
reducibility of the $\kappa$--symmetry and makes it possible to carry 
out covariant  Hamiltonian analysis of superstring dynamics at least on 
the classical level);

\noindent
ii) the geometrical ground for this is that the theory of super--p--branes is
supposed to be a particular kind of doubly supersymmetric models (studied
earlier in \cite{doubly}) which describe an embedding of supersurfaces into
target superspaces. (This naturally incorporates twistors into the theory).

One of the aims of the approach has been to push forward the
problem of covariant quantization of superstring theory. As a result the
methods to attain this objective have undergone substantial modifications
during last few years (see \cite{berk} for a recent review).

At the same time new directions have been revealed where the doubly
supersymmetric formulation can be used. For instance,
it can serve as a natural
dynamical basis for generalizing geometrical
methods and notions of classical
surface theory (which have been used in
bosonic string theory \cite{geo}) to
study the embedding of at least a particular class of supersurfaces
corresponding to super--p--branes, and then to apply the geometrical methods
back to the analysis of variety 
of fundamental and solitonic super--p--branes we are having at hand. 
(Supergravity as a theory of supersurfaces was considered previously in 
\cite{rs}).

Recent results in this direction \cite{bpstv} have revealed  links
of Green--Schwarz superstring dynamics with a nonstandard version of the
super--Liouville system \cite{lio} and with super--WZNW models subject to a
nonstandard Hamiltonian reduction \cite{wz}. Through this WZNW connection
we have arrived at a superconformal model related to a universal
string theory describing a hierarchy of superstrings embedded one into
another \cite{ber}.

The main part of this report is devoted to the discussion of these points
by use of the example of a Green--Schwarz superstring propagating in $N=2$,
$D=3$ flat superspace.

We shall see how by specifying the embedding of
worldsheet supersurface swept by the superstring 
one can solve for the
Virasoro constraints and reduce 
superstring equations of motion to a
supersymmetric Liouville--like system of equations.

In the doubly supersymmetric formulation worldsheet of the Green--Schwarz
superstrings is a supersurface parametrized by two bosonic coordinates
$\xi^m=(\tau+\sigma,\tau-\sigma)$, $m=(++,--)$ 
and fermionic coordinates $\eta$ whose
number to be equal to the number of independent $\kappa$--symmetry
transformations in the standard Green--Schwarz formulation. In our case there
are one left-- and one right--handed Majorana--Weyl spinor coordinates, which
means that we deal with $n=(1,1)$ local supersymmetry on worldsheet
supersurface
\begin{equation}\label{ws}
{\cal M}_{ws}:~~~Z=(\xi^m,~\eta^+,~\eta^-),
\end{equation}
where $+$ and $-$ stand for light-cone spinor indices of $SO(1,1)$.

To describe ${\cal M}_{ws}$ geometry one should set on ${\cal M}_{ws}$
a local frame of supervielbein one--forms which contains two bosonic vector
and to fermionic spinor components
\begin{equation}\label{e}
e^A(Z)=\left(e^a(Z), ~e^+(Z),~e^-(Z)\right), ~~~a=(++,--).
\end{equation}

We consider an embedding of ${\cal M}_{ws}$ into  $N=2$, $D=3$ flat
superspace--time paramet\-rized 
by three bosonic vector and two Majorana spinor
coordinates
$X^{\underline m}(Z),~\Theta^{1\underline \mu}(Z),$
$\Theta^{2\underline \mu}(Z),$
where ${\underline m}=0,1,2$ and ${\underline \mu}=1,2$ are
vector and spinor indices of the $D=3$ Lorentz group $SO(1,2)$, respectively.

A natural supersymmetric rigid frame in flat target superspace is 
\begin{equation}\label{rf}
\Pi^{\underline m}=dX^{\underline m}-i\bar\Theta^{i}
\Gamma^{\underline m}d\Theta^{i}, \qquad d\Theta^{i\underline \mu} \qquad
(i=1,2).
\end{equation}
$\Gamma^{\underline m}_{\underline{\a\b}}$ are $D=3$ Dirac matrices.

The study of ${\cal M}_{ws}$ embedding is started with fitting the rigid
target--superspace frame \p{rf} to that on the supersurface \p{e}.
To this end we transform \p{rf} into a new local frame 
\begin{equation}\label{lf}
E^{\underline a}=\Pi^{\underline m}
u_{\underline m}^{\underline a}(X,\Theta),\qquad
E^{i \underline \alpha}=d\Theta^{i\underline \mu}v_{\underline \mu}^
{\underline \alpha}(X,\Theta),
\end{equation}
where $u_{\underline m}^{\underline a}$ and $v_{\underline\mu}^
{\underline \alpha}$ are matrices of the vector and spinor
representation of the target--superspace Lorentz group $SO(1,2)$, 
respectively. But since the vector and spinor
components of \p{rf} are subject to the Lorentz transformation simultaneously,
$u$ and $v$ are not independent and connected through 
the well--know twistor--like relation
$
u_{\underline m}^{\underline a}(\Gamma^{\underline m})_{\underline
{\mu\nu}}= v^{\underline \alpha}_{\underline \mu}
(\Gamma^{\underline a})_{\underline{\alpha\beta}}
v^{\underline \beta}_{\underline \nu}
$
between vectors and commuting spinors.
This explains why the approach is called ``twistor--like".

From the analysis of superstring dynamics in the twistor--like approach we
learn \cite{bpstv} 
that the target--superspace local frame \p{lf} can be attached
to the supersurface as follows:
\begin{equation}\label{pull}
E^{\perp}(Z) = 0, \qquad
E^a(Z) = \Pi^{\underline m}u^a_{\underline m}=
(dX^{\underline m}-i\bar\Theta^{i}
\Gamma^{\underline m}\Theta^{i})u^a_{\underline m}=e^a,
\end{equation}
\begin{equation}\label{pullb}
E^{1+}(Z) = d\Theta^{1\underline \mu}v^{+}_{\underline \mu}=e^+,\qquad
E^{2-}(Z) = d\Theta^{2\underline \mu}v^{-}_{\underline \mu}=e^-,
\end{equation}
where the target superspace indices split onto that of ${\cal M}_{ws}$ 
and of the orthogonal vector direction 
(${\underline a}\rightarrow (a,\perp);~ {\underline \alpha}\rightarrow 
(+,-))$.

Eq. \p{pull} tells us that one of the vector components of
the target superspace frame can be made orthogonal to the supersurface
(its pullback on ${\cal M}_{ws}$ is zero) and three other
relations in \p{pull}, \p{pullb} identify (on ${\cal M}_{ws}$) 
components of the target superspace
frame with the intrinsic ${\cal M}_{ws}$ supervielbein components \p{e}.

From \p{pull}, \p{pullb}, using the orthogonality properties
of $u_{\underline m}^{\underline a}$, we find that the pullback on
${\cal M}_{ws}$  of the vector one--superform
$
\Pi^{\underline m}(Z)=dX^{\underline m}-i\bar\Theta^{i}
\Gamma^{\underline m}d\Theta^{i}
=e^au_a^{\underline m}(Z)
$
is zero along the
fermionic directions $e^{\pm}$ of ${\cal M}_{ws}$:
\begin{equation}\label{gdc}
D_{\pm}X^{\underline m}-i\bar\Theta^{i}
\Gamma^{\underline m}D_{\pm}\Theta^{i}=0,
\end{equation}
where $D_{\pm}$ are covariant spinor derivatives on ${\cal M}_{ws}$.
Eq. \p{gdc} is called the geometrodynamical condition. 
Eqs. \p{pull} guarantee that the
Virasoro constraints on superstring dynamics
\begin{equation}\label{vc}
\Pi^{\underline m}_{++}\Pi^{\underline m}_{++}=0=
\Pi^{\underline m}_{--}\Pi^{\underline m}_{--}
\end{equation}
are identically satisfied for such kind of embedding \cite{bpstv}. 

One can notice that Eqs. \p{pull}, \p{pullb} 
are first--order differential
equations on $X(Z)$ and $\Theta(Z)$. If they are solved one would know the
shape of the worldsheet supersurface in $D=3$, $N=2$ superspace and thus 
would
solve the problem of describing classical superstring motion.
To solve \p{pull} and \p{pullb} 
one must know $v^{\underline \alpha}_{\underline \mu}(Z)$
and the components of $e^a(Z), e^{\pm}(Z)$. To get this information one
should study the integrability conditions of 
\p{pull}, \p{pullb} 
which are obtained by taking the external differential
of \p{pull}, \p{pullb}. Basic integrability conditions thus obtained are:
\begin{equation}\label{con}
de^a-\Omega^a_b e^b=ie^\a\gamma^a_{\a\b}e^{\b}=T^a,
\end{equation}
\begin{equation}\label{sec}
\Om^{\perp a}=K^a_b e^b+K^a_\a e^\a,
\end{equation}
where $\a=(+,-)$, $\gamma^a_{\a\b}$ are d=2 Dirac matrices, and external
differentiation and product of the forms are implied.
Eqs. \p{con}, \p{sec} contain one forms $\Omega^a_b, \Om^{\perp a}$ which are
Cartan forms of the $SO(1,2)$ Lorentz group constructed out of the matrix
$v^{\underline \alpha}_{\underline \mu}$ components:
\begin{equation}\label{cartan}
\Om^{ab}=\e^{ab}v^{+}_{\underline \mu}dv^{-\underline \mu},
\qquad
\Om^{\perp a}=\gamma^a_{\a\b}v^{\alpha}_{\underline \mu}
dv^{\b\underline \mu}.
\end{equation}

Eq. \p{con} determines parallel transport of vector supervielbeins along
 ${\cal M}_{ws}$ carried out by induced connection $\Om^{ab}$. It reads
that the connection possesses torsion whose spinor--spinor components are
constrained to be equal to the $\g$--matrix components. This is a basic
torsion constraint of any supergravity theory. In the
geometrical approach it is not imposed by hand, but appears as a
consistency condition of ${\cal M}_{ws}$ embedding.

The second condition \p{sec} specifies the expansion of $\Om^{\perp a}$ in
the ${\cal M}_{ws}$ supervielbein components: the bosonic matrix
$K_{ab}(Z)$ is symmetric and has the properties of the second fundamental
form analogous to that of the bosonic surfaces, while spinor components
$K^\a(Z)\equiv K^a_{\b}\gamma_{a}^{\b\a}$ of the Grassmann--odd spin--tensor
$K^a_{\b}$ can be associated with a fermionic counterpart of the second
fundamental form along Grassmann directions of the supersurface.

By construction the Cartan forms \p{cartan} must satisfy the $SO(1,2)$
Maurer--Cartan equations $d\Om-\Om\Om=0$ which split into two systems of
equations with respect to the world--sheet indices:
\begin{equation}\label{cod}
d\Om^{\perp a}-\Om^a_b\Om^{\perp b}=0,
\end{equation}
\begin{equation}\label{gauss}
R^{ab}=d\Om^{ab}=\Om^{\perp a}\Om^{\perp b}.
\end{equation}
Eq. \p{cod} is known as the Codazzi 
equation and \p{gauss} is called the Gauss
equation in surface theory. On the other hand one can recognize in \p{cod}
and \p{gauss} relations which specify geometry on a two--dimensional coset
space ${SO(1,2)}\over{SO(1,1)}$ with $\Om^{\perp a}$ being a vielbein,
$\Om^{ab}$ being a spin connection and $R^{ab}$ being a constant curvature
tensor of  ${SO(1,2)}\over{SO(1,1)}$.

Thus we have reduced the problem of studying superstring dynamics (as
the embedding of a supersurface into target superspace) 
to study a mapping
of ${\cal M}_{ws}$ onto the bosonic coset space of constant curvature.
It is here that a connection of Green--Schwarz superstring dynamics with an
$n=(1,1)$ super--WZNW model comes out.

To completely describe superstring dynamics in geometrical terms we should
specify what additional conditions on embedding arise when the superstring
equations of motion are taken into account. In bosonic surface theory
such an embedding is called minimal and is characterized by traceless second
fundamental form $K_{ab}$ :
\begin{equation}\label{tl}
K^a_a=0.
\end{equation}
In the supersymmetric case we get analogous condition on the bosonic part of
the second fundamental form \p{sec} which is in one to one correspondence
with $X^{\underline m}$ equations of motion
$
D_{a}\left(D^{a}X^{\underline m}-i\bar\Theta^{i}
\Gamma^{\underline m}D^{a}\Theta^{i}\right)u_m^{\perp}=0
$
(where $D_a=(D_{--},D_{++})$ is a vector covariant derivative). In
addition, $\Theta(z)$ equations of motion, which in the twistor--like
approach have the form
$
D_{--}\Th^{1\underline \mu}v^-_{\underline \mu}=0,
$
$
D_{++}\Th^{2\underline \mu}v^+_{\underline \mu}=0,
$
result in vanishing the fermionic part of the second fundamental form
in Eq. \p{sec}, namely
\begin{equation}\label{f}
K^\a(Z)\equiv K^a_{\b}\gamma_{a\b}^\a=0.
\end{equation}

The minimal embedding conditions \p{tl}, \p{f} further reduce the number
of independent superfields which determine the induced geometry on
${\cal M}_{ws}$. One can show that in a superconformal gauge
$\Om^{\perp a}$ and $\Om^{ab}$ (which bear all information about
${\cal M}_{ws}$) depend only on three superfields, one bosonic superfield
$\Phi(Z)$ and two fermionic superfields $\Psi_+(Z)$ and $\bar\Psi_-(Z)$,
whose leading components describe one bosonic and two fermionic 
physical degrees of freedom of the classical N=2, D=3 Green--Schwarz
superstring.

The superfields obey $n=(1,1)$ superconformal invariant equations which
follow from \p{cod} and \p{gauss}.
\begin{equation}\label{chi}
D_+\Psi_+=0, \qquad  D_-\bar\Psi_-=0,
\qquad {\rm (anti)chirality~ conditions}
\end{equation}
\begin{equation}\label{lio}
D_+D_-\Phi=e^{2\Phi}\bar\Psi_-\Psi_+\qquad 
~~~~~~~{\rm Liouville-like~ equation}
\end{equation}
\begin{equation}\label{const}
D_-\Psi_++2D_-\Phi\Psi_+=1,\qquad
D_+\bar\Psi_-+2D_+\Phi\bar\Psi_-=1,\qquad
{\rm ``constraints"}.
\end{equation}
Let us compare the system \p{chi}--\p{const} with the standard
super--Liouville equation 
\begin{equation}\label{li}
D_+D_-\tilde\Phi=-ie^{\tilde\Phi},
\end{equation}
where $\tilde\Phi(Z)=\tilde\phi+i\eta^+\tilde\psi_++i\eta^-\tilde\psi_-
+i\eta^+\eta^-F$ is an unconstrained bosonic superfield and 
$D_-,D_+$ are covariant spinor derivatives 
$(D_-^2=i\partial_{--},~D_+^2=i\partial_{++})$.

Recently Ivanov, Krivonos and Pashnev found relations (local B\"acklund 
transformations) which express the
component fields of the standard super--Liouville system in terms of 
the components of the alternative one. 
\footnote
{However, it seems not possible to locally express the fields of
\p{chi}--\p{const} in terms of $\tilde\Phi(Z)$ 
components because of derivatives
in the r.h.s. of \p{rel}.} 
They can be packed into the following
superfield expression, which has the form of expansion of $\tilde\Phi(Z)$
in series of $\bar\Psi_-(Z), \Psi_+(Z)$:
\begin{equation}\label{rel}
\tilde\Phi(Z)=\Phi(Z)+D_+\Phi\bar\Psi_-(Z)+D_- \Phi\Psi_+(Z)+
ie^{\Phi}\bar\Psi_-(Z)\Psi_+(Z).
\end{equation}
In spite of this relation two super--Liouville systems possess different
properties. For instance, in components \p{chi}--\p{const} reduce to the
purely bosonic Liouville equation and two free chiral fermion equations
\begin{equation}\label{comp}
\partial_{++}\partial_{--}\phi(\xi)=-e^{2\phi},\qquad
\partial_{++}\psi_+=0=\partial_{--}\psi_-,
\end{equation}
while the standard super--Liouville system \p{li}
is characterized by nontrivial
coupling between the boson and the fermions
\begin{equation}\label{kulish}
\partial_{++}\partial_{--}\tilde\phi=
-e^{\tilde\phi}(e^{\tilde\phi}+i\tilde\psi_-\tilde\psi_+),
\qquad\partial_{++}\tilde\psi_-=- e^{\tilde\phi}\tilde\psi_+, \qquad
\partial_{--}\tilde\psi_+=e^{\tilde\phi}\tilde\psi_-.
\end{equation}
The supersymmetry transformation properties
of $\psi_+$, $\psi_-$ 
are ($\delta\eta_-=\e_-(\xi^{++})$,  $\delta\eta_+=\e_+(\xi^{--})$):
\begin{equation}\label{susy}
\delta\psi_+=\e_+(1+i\psi_+\partial_{--}\psi_+), \qquad
\delta\psi_-=\e_-(1+i\psi_-\partial_{++}\psi_-). \qquad
\end{equation}
We see that the fermionic fields transform as Goldstone fermions 
\cite{av} 
which points to the spontaneous breaking of the superconformal symmetry in
contrast to the standard super--Liouville system.

Another difference from the standard super--Liouville fermions is that
$\psi_+$ and $\psi_-$ have conformal spin $(-{1\over 2})$ (unusual for
matter fields), while the former have conformal spin $1\over 2$.

In \cite{wz} it was shown that $\psi_+$ and $\psi_-$  together with 
their momenta belong to a ghost--like
$b-c$ system of conformal spin $({3\over 2},-{1\over 2})$ which one
encounters with in universal string theory \cite{ber}. It seems of interest that
in our case such a system arose as one describing the classical physical
degrees of freedom of a conventional $N=2$, $D=3$ Green--Schwarz superstring.

The reason why the alternative and not the standard super--Liouville system 
of equations was obtained in the case of the free 
$N=2$, $D=3$ Green--Schwarz superstring is because of their another 
difference: the underlying group of the standard $n=(1,1)$ super--Liouville 
equation is the supergroup $OSp(1|2)$ while that of the alternative system
is $SL(2,{\bf R})\sim SO(1,2)$, as we have seen above. But since there is 
the relation \p{rel} between the two systems one may argue that on the mass 
shell the $N=2$, $D=3$ Green--Schwarz superstring possesses hidden 
$OSp(1|2)$ symmetry.

By the statement about the underlying groups I mean that the standard 
super--Liouville model can be obtained by a Hamiltonian reduction of an
$n=(1,1)$ super--WZNW model based on  $OSp(1|2)$ \cite{drs}, while the 
system \p{chi}--\p{const} arises in an appropriately constrained super--WZNW 
model based on $SL(2,{\bf R})$ \cite{wz}. Let us briefly sketch this point.

The equations of motion of the super--WZNW models \cite{wznw,drs} are 
(anti)chirality conditions on fermionic supercurrents taking their 
values in the algebra of a (super)group $G(Z)$:
\begin{equation}\label{chira}
D_+({1\over i}D_-GG^{-1})\equiv D_+\hat\Psi_-=0,
\qquad
D_-({1\over i}D_+G^{-1}G)\equiv D_-\hat{\bar\Psi}_+=0.
\end{equation}
We see that $\hat\Psi_-(Z)$ and $\hat{\bar\Psi}_+(Z)$ are the fermionic 
components of Cartan forms on $G$ (like in the consideration above).
When $G=OSp(1\vert 2)$ they have the following expansion in $OSp(1\vert 
2)$ generators:
$$
\hat\Psi_-=\Psi_-H+\Psi_+E_{--}+\Psi_{---}E_{++}+\Psi F_-+\Psi_{--}F_+,
$$
\begin{equation}\label{ex}
\hat{\bar\Psi}_+=\bar\Psi_+H+\bar\Psi_{+++}E_{--}+\bar\Psi_{-}E_{++}
+\bar\Psi_{++}F_-+\bar\Psi F_+,
\end{equation}
where $H,E_{++}, E_{--}$ are the bosonic generators of the $Sl(2,{\bf 
R})$ 
subgroup and $F_+,F_-$ are fermionic generators of $OSp(1\vert 2)$.
Pluses and minuses stand for charges of the quantities with respect to 
the Cartan subalgebra $H$ of $OSp(1\vert 2)$.

To get super--Liouville equations from \p{chira}, one should first identify a 
superfield $\Phi(Z)$ which will describe the Liouville modes. For that 
one takes the Gauss decomposition of the group element
\begin{equation}\label{gd}
G=g_+e^{\Phi H}g_-,
\end{equation}
where $g_+(Z)$ and $g_-(Z)$ are generated by $(E_{++},F_+)$ and 
$(E_{--},F_-)$, respectively.

Then one should perform a Hamiltonian reduction procedure \cite{f}
by imposing constraints on components of \p{ex} as 
follows:
\begin{equation}\label{hr}
\bar\Psi=1,~~~ \bar\Psi_-=0;\qquad \Psi=1,~~~ \Psi_+=0.
\end{equation}

Note that $\Psi,\bar\Psi$ are bosonic superfields (see \p{ex}), so they 
can be constrained to be nonzero constants, while $\Psi_+,\bar\Psi_-$ 
are fermionic superfields.
Such a reduction of supercurrent components \p{ex} and Eqs. \p{chira} 
results in the standard super--Liouville equation \p{kulish}.

If $G=Sl(2,{\bf R})$ then all supercurrents corresponding to $F_+,F_-$ 
in \p{ex} are identically zero and we remain only with  
$Sl(2,{\bf R})$ supercurrent components which are fermionic.

So, it turns out that now, to get the alternative super--Liouville 
system \p{chi}--\p{const} one should first construct bosonic 
supercurrents out of the fermionic ones using the Maurer--Cartan 
equations on the superforms $dGG^{-1}$ and $G^{-1}dG$ and then
impose constraints on their $E_{++},E_{--}$ components \cite{wz}:
$$
J\equiv tr(\partial_{--}GG^{-1}E_{++})=D_-\Psi_++2i\Psi_-\Psi_+=1,
$$
\begin{equation}\label{j}
\bar J\equiv 
tr(G^{-1}\partial_{++}GE_{--})=D_+\bar\Psi_-+2i\bar\Psi_+\bar\Psi_-=1,
\end{equation}
In \p{j} one can recognize \p{const}, while \p{chi} and \p{lio} follow 
from \p{chira}.

To conclude, let us sum up what we have learned from the 
consideration above.

The doubly supersymmetric formulation of super--p--branes 
allows one to generalize methods of surface theory, to apply the 
geometrical approach to study the dynamics of supersymmetric extended 
objects and to reduce their equations of motion to nonlinear 
supersymmetric systems of equations on their physical modes.
such as the alternative version of the super--Liouville
system in the case of the $N=2$, $D=3$ Green--Schwarz superstring.
The generalization of these results to higher--dimensional and curved 
supergravity backgrounds might  be useful for studying problems of 
superstring cosmology (see \cite{ls} and refs. therein), and 
super--p--branes in superstring theory, M-- and 
F--theory.

We have also observed a connection of superstring dynamics with 
a constrained super--WZNW model. This is in accordance with a 
well known fact that WZNW models have deep relationship with string theory.
A new point we encountered with is that to constrain the $n=1$ 
$Sl(2,{\bf R})$ super--WZNW model in the appropriate way we had to 
impose the nonlinear 
constraints \p{j} on basic fermionic supercurrents, which turn out to be 
a mixture of first-- and second--class constraints \cite{wz}. Usually 
one deals with linear first--class constraints like Eqs. \p{hr}. Such a 
generalization of the Hamiltonian reduction procedure allows one to 
involve into the game wider class of super--WZNW models and 
corresponding affine (super)algebras which contain bosonic simple roots 
in any root decomposition as, for example, $OSp(1\vert 4)$.
This way one may hope to get new supersymmetric versions of Toda--like 
systems with nonlinearly realized supersymmetry and nonstandard 
realizations of corresponding $W$--algebras.

\medskip
This work was partially supported   
by the State  Committee  for  Science  and  Technology  of
Ukraine under the Grant N 2/100 and
 by the INTAS grant 93--493.

\end{document}